\documentclass[smallextended]{svjour3}
\smartqed
\usepackage{graphicx}
\usepackage{mathptmx}
\usepackage{latexsym}
\usepackage{xcolor}
\journalname{arXiv}
\RequirePackage{fix-cm}
\begin{document}
\title{On the BTZ Black Hole and the Spinning Cosmic String}
\author{Reinoud Jan Slagter \and  Jebin Larosh}
\institute{R. J. Slagter \at  ASFYON, Astronomisch Fysisch Onderzoek Nederland  and \\University of Amsterdam, Department of Physics, The Netherlands. \\ \email{info@asfyon.com}
\and
J. Larosh \at Manonmaniam Sundaranar University, Department of Physics  \\ Tamilnadu, India \\ \email{jawan.jebin@gmail.com}}
\date{Received: 18 Dec 2019}
\maketitle
\begin{abstract}
We review the Ba\u nados-Teitelboim-Zanelli (BTZ) black hole solution in connection with the spinning string solution. We find a new exact solution, which can be related to the $(2+1)$-dimensional spinning point particle solution. There is no need for a cosmological constant, so the solution can be up-lifted to $(3+1)$ dimensions.
The exact solution in a conformal invariant gravity model, where the spacetime is written as $g_{\mu\nu}=\omega^2 \tilde g_{\mu\nu}$, is horizon free and has an ergo-circle, while $\tilde g_{\mu\nu}$ is the BTZ solution. The dilaton $\omega$ determines the scale of the model. In accordance with the spinning cosmic string solution, it is conjectured that the new solution can be linked to the mass of the interior of the spinning cosmic string.
\end{abstract}
\section{Introduction}
Besides the well-studied Schwarzschild and Kerr solution in general relativity theory (GRT), there is another black hole solution in $(2+1)$-dimensional spacetimes, i.e., the Ba\u nados-Teitelboim-Zanelli (BTZ) black hole\cite{banadoz:1993}. The BTZ geometry solves Einstein's equations  with a negative cosmological constant in $(2+1)$-dimensions. The solution asymptotes for large $r$ to a global anti-de Sitter ($AdS_3$) spacetime.
In general, $(2+1)$-dimensional gravity has been widely recognized as an interesting laboratory not only for studying GRT, but also quantum-gravity models\cite{compere:2019}.
It is conjectured that this genuine solution will be of importance  when one considers thermodynamic properties close to the horizon, i.e., Hawking radiation\cite{thooft:2017}.
The BTZ solution is comparable with the spinning point particle solution ("cosmon"\cite{deser:1984,deser:1989,deser:1992}) of the dimensional reduced spinning cosmic string or Kerr solution.
$(2+1)$-dimensional gravity without matter, implies that the Ricci- and Riemann tensor vanish, so matter-free regions are flat pieces of spacetime. When locally a mass at rest  is present, it cuts out a wedge from the 2-dimensional space surrounding it and makes the space conical. The angle deficit is then proportional to the mass\cite{garfinkle:1985}.
The important fact is that the spinning point particle has a physical acceptable counterpart in $(3+1)$-dimensions, i.e., the spinning cosmic string. The z-coordinate is suppressed, because there is no structure
in that direction altogether.
It is the unconventional range and jump properties of the coordinates that remind us that there are sources somewhere.
One can proof that the source of the $(3+1)$-dimensional cosmic string cannot be infinite thin\cite{geroch:1987}. This implies matching problems at the boundary\cite{janca:2007,krisch:2003}.
This problem could be overcome in a conformal invariant setting\cite{slagter:2019}.
The BTZ solution, however, when up-lifted, needs a zero cosmological constant, so a different solution emerge.
It is not a surprise that these models are used in constructing quantum gravity models. In these models one uses locally Minkowski spacetime, so planar gravity fits in very well.
It is conjectured\cite{thooft:1996} that $(2+1)$-dimensional gravity with matter could be quantized in a unambiguous way.
In this context, one can also consider conformal invariant (CI) gravity models, specially after the recognition that the asymptotically $AdS_3$ is related to a 2-dimensional conformal field theory (CFT)\cite{strom:1997}.
Conformal invariance was originally introduced by Weyl\cite{weyl:1918}. The idea was to introduce a new kind of geometry, in relation to a unified theory of gravitation and electromagnetism. This approach was later abandoned with the birth of modern gauge field theories.
Quite recently the Anti-deSitter/Conformal field-theory (AdS/CFT)   correspondence renewed the interest in conformal gravity. AdS/CFT is a conjectured relationship between two kinds of physical theories. AdS spaces  are used in theories of quantum gravity while  CFT  includes theories similar to the Yang–Mills theories that describe elementary particles\cite{mald:1998}.
It is now believed that CI can help us move a little further along the road to quantum gravity.
CI in GRT considered as exact at the level of the Lagrangian but spontaneously broken just as in the case of the Brout-Englert-Higgs mechanism (BEH) in standard model of particle physics, is an approved alternative for disclosing the small-distance structure when one tries to describe quantum-gravity problems\cite{thooft:2011}.  It  can also be used to model scale-invariance in the  cosmic microwave background radiation (CMBR)\cite{bars:2014}. Another interesting application  can be found in the work of Mannheim on conformal cosmology\cite{mannheim:2005}. This model could serve as an alternative approach to explain  the rotational curves of galaxies, without recourse to dark matter and dark energy (or cosmological constant).
The key problem is the handling of asymptotic flatness of isolated systems in GRT, specially when they radiate and the generation of the metric $g_{\mu\nu}$ from at least Ricci-flat spacetimes.
In the non-vacuum case one should construct a Lagrangian where spacetime and the fields defined on it, are topological regular and physical acceptable. This can be done by considering the scale factor( or warp factor in higher-dimensional models) as a dilaton field besides, for example, a conformally coupled  scalar field or other fields.
Conformal invariant gravity distinguishes itself by the notion that the spacetime is written as $g_{\mu\nu}=\omega^2 \tilde g_{\mu\nu}$, with $\omega$ a dilaton field which contains all the scale dependencies  and $\tilde g_{\mu\nu}$ the "un-physical" spacetime, related to the $(2+1)$-dimensional Kerr and BTZ black hole solution.

In this manuscript we present a new solution of the BTZ-type and compare the solution with the conformal invariant counterpart solution. We will not consider here, for the time being, the quantum mechanical implications of the model.
\section{The BTZ solution revised}
If one solves the Einstein equations $G_{\mu\nu}=\lambda g_{\mu\nu}$ for the spacetime
\begin{eqnarray}
ds^2=-N(r)^2 dt^2+\frac{1}{N(r)^2}dr^2+r^2\Bigl(d\varphi+N^\varphi(r)dt\Bigr)^2,\label{1}
\end{eqnarray}
one obtains
\begin{eqnarray}
N(r)^2\equiv \alpha^2-\lambda r^2+\frac{16 G^2J^2}{r^2}  , \qquad  N^\varphi(r)\equiv -\frac{4GJ}{r^2} +S\label{2},
\end{eqnarray}
where $S$, $J$ and $\alpha$ are integration constants\cite{banadoz:1993,compere:2019}.  The parameters $\alpha$ and $J$  represent the standard ADM mass ($\alpha^2 =\pm 8GM$) and angular momentum and determine the asymptotic behavior of the solution. $\lambda$ represents the cosmological constant. There is an inner and outer horizon  and an ergo-circle just as in the case of the Kerr spacetime. However, if one lifts-up this spacetime to $(3+1)$ dimensions, one must take $\lambda=0$, which can easily be verified by the Einstein equations. So we consider here the case $\lambda=0$, and we write the spacetime as
\begin{equation}
ds^2=-\Bigl[8G(JS-M)-S^2r^2\Bigr]dt^2+\frac{r^2r_H^2}{16G^2J^2(r_H^2-r^2)}dr^2+r^2d\varphi^2+2r^2\Bigl(S-\frac{4GJ}{r^2}\Bigr)dt d\varphi\label{3},
\end{equation}
with $r_H$ the horizon $r_H=\sqrt{\frac{2G}{M}}J$.
In the case of $S=0$, which is also done  in the original BTZ solution, one can transform the spacetime to
\begin{equation}
ds^2=-\Bigl(\alpha dt+\frac{4GJ}{\alpha}d\varphi \Bigr)^2+dr'^2+\alpha^2 r'^2 d\varphi^2 \label{4}
\end{equation}
by $r'^2=\frac{16G^2J^2+\alpha^2 r^2}{\alpha^4}$.
This is just the spinning particle spacetime\cite{deser:1989}.
One is left with cosmological spacetimes without horizons.  There are now evidently CTC's for $r'<\frac{4GJ}{\alpha^2}$. By re-defining $\varphi \rightarrow \alpha\varphi$, in order to obtain asymptotically the Minkowski spacetime, we can identify the mass parameter $\alpha$ with  an angle deficit (ignoring for the moment the time re-definition in order  to get rid of the $J$).
Already mentioned before, we then run into problems (apart from the zero cosmological constant) when lifting-up to $(3+1)$-spacetime.
The mass of the general relativistic cosmic string is determined by the parameters of scalar-gauge fields, i.e., by taking the integral of the energy density over a t=constant, z=constant two-surface. So there is an obscurity by defining the mass parameter M of the bounded (2+1)-dimensional BTZ black hole by using the method of the surface charges associated to the 2 Killing vectors\cite{compere:2019}.
One must keep in mind that the original asymptotic solution of the general relativistic cosmic string is given by\cite{garfinkle:1985}
\begin{equation}
ds^2=-e^{a_0}(dt^2-dz^2)+dr^2+e^{-2a_0}(kr+a_1)^2d\varphi^2\label{5},
\end{equation}
with $a_0$ and $a_1$ integration constant, but $k$ follows directly from the field equations. The transformations $ r \rightarrow r+\frac{a_1}{k}, \varphi \rightarrow ke^{-a_0}\varphi, z \rightarrow e^{-a_0/2}z, t \rightarrow e^{-a_0/2}t$ will now deliver Minkowski minus a wedge, i.e., a  conical spacetime.
\begin{equation}
ds^2=-dt'^2+dz'^2+dr'^2+(1-4G\mu)^2r'^2d\varphi^2\label{6},
\end{equation}
where we wrote  $ 1-4G\mu =k e^{-a_0}  $. The angle deficit is then given by
the parameters $k$ and $a_0$ and are in general determined by the string variables, i.e., the scalar and gauge field ( and by the metric variables due to the fact that we are dealing with a coupled set of differential equations). No a priori connection with the mass of a black hole is necessary: the mass density of the cosmic string is directly related to the angle deficit when lifting-up to 4D.

\subsection{The new solution}
For $S=\frac{2M}{J}$ we can write the spacetime of Eq.(\ref{3}) for the transformation
\begin{equation}
r^*=-\frac{r}{8GM}+\frac{\sqrt{2}J}{8M\sqrt{GM}} arctanh\Bigl(\frac{r}{r_H}\Bigr)\label{7},
\end{equation}
in the form
\begin{equation}
ds^2=16G^2J^2\Bigl(\frac{1}{r^2}-\frac{1}{r_H^2}\Bigr)(-dt^2+dr^{*2}) +r^2\Bigl[d\varphi +4GJ\Bigl(\frac{1}{r_H^2}-\frac{1}{r^2}\Bigr)dt\Bigr]^2\label{8}.
\end{equation}
The angular velocity of the null generator, $\Omega_A\equiv\frac{d\varphi}{dt}=\frac{g_{t\varphi}}{g_{\varphi\varphi}}$ becomes
\begin{equation}
\Omega_A=4GJ\Bigl(\frac{1}{r_H^2}-\frac{1}{r^2}\Bigr)\label{9}.
\end{equation}
\begin{figure}[h]
\centerline{
\includegraphics[width=6cm]{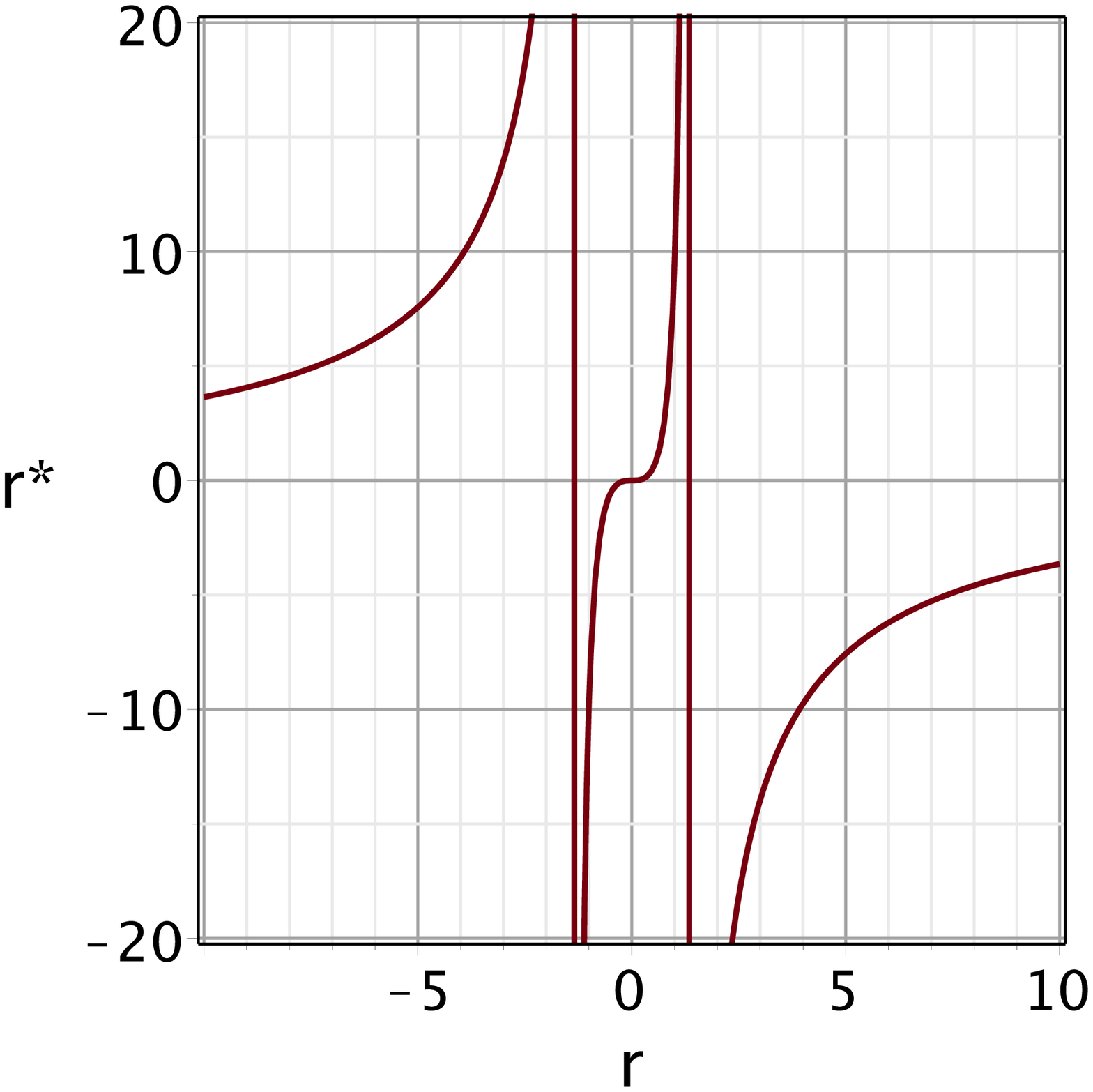}
\includegraphics[width=6cm]{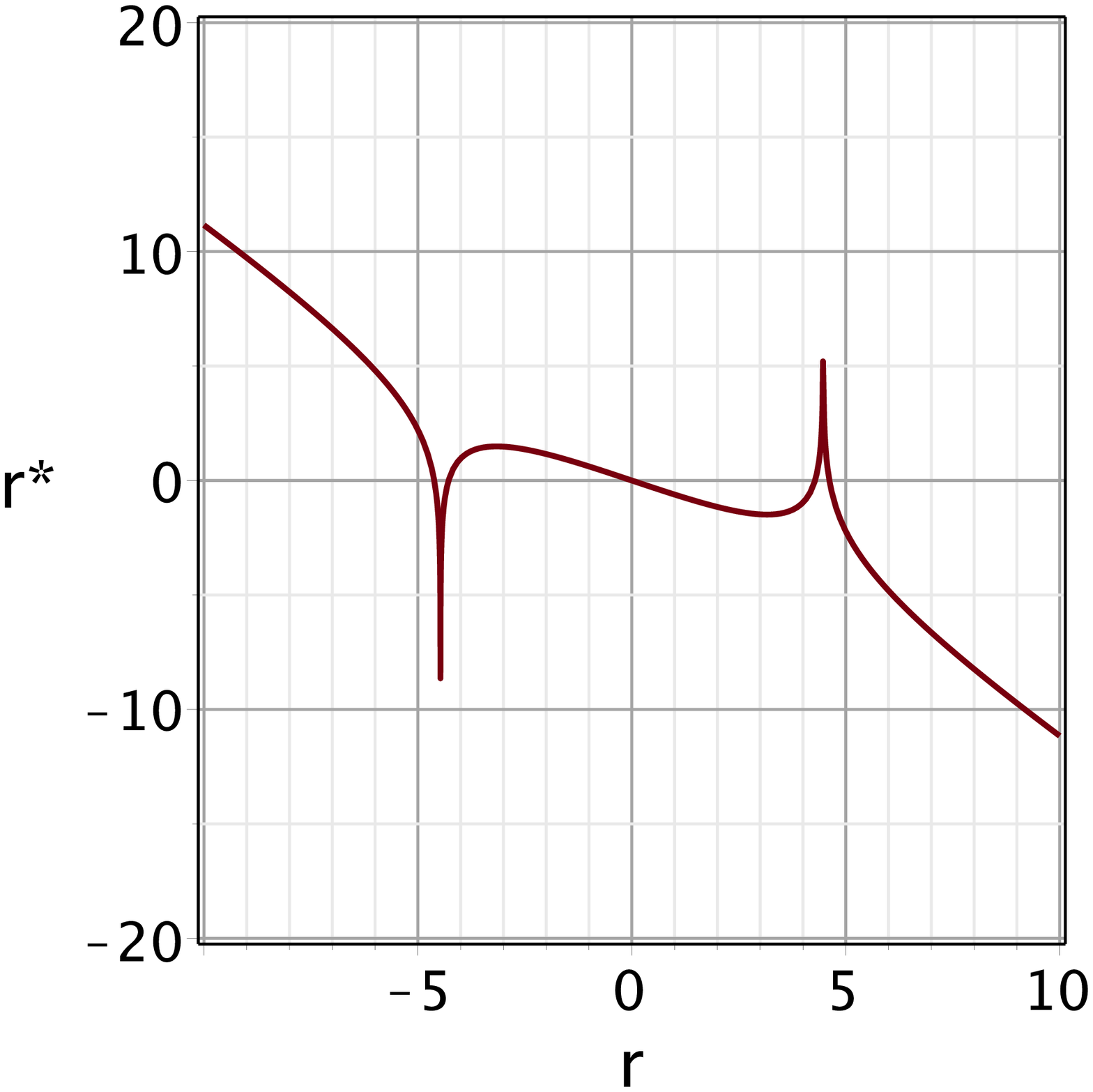}}
\vspace{1cm}
\caption{Graphs of $r^*$ for the original BTZ solution (left) and our solution of Eq.(\ref{7}) (right).  }
\end{figure}
The $dtd\varphi$ term will then be zero at the horizon, which means that locally non-rotating observers has no coordinate angular velocity, i.e., there is no dragging of inertial frames at the horizon.
We can plot the Penrose diagram by defining the appropriate null coordinates $U=r^*-t, V=r^*+t$  and switch to compactified coordinates $ U=tan\Bigl(\frac{p+q}{2}\Bigr), V=tan\Bigl(\frac{p-q}{2}\Bigr)$.
So $r^*=\frac{sin(p)}{cos(p)+cos(q)}$ as in the original BTZ case.
The metric then becomes
\begin{equation}
ds^2=\frac{16G^2J^2\Bigl(\frac{1}{r^2}-\frac{1}{r_H^2}\Bigr)}{(cos(p)+cos(q))^2}\bigr[ dp^2-dq^2\Bigr]+r^2\Bigl[d\varphi +4GJ\Bigl(\frac{1}{r_H^2}-\frac{1}{r^2}\Bigr)dt\Bigr]^2\label{10}.
\end{equation}
In figure 1 we plotted $r^*$ for the original BTZ solution in 3D and our solution of Eq.(\ref{7}). The only difference is the behavior of $r^*$ at $r=r_H$: there is, in our solution, no jump in $r^*$ when crossing the horizon. Further, for $r\rightarrow \pm \infty$, we have $r^* \rightarrow \mp \infty$.
\begin{figure}[h]
\centerline{
\includegraphics[width=5cm]{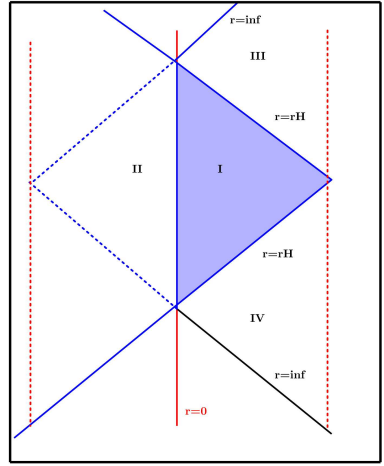}}
\caption{Penrose diagram of the new solution in $(p,q)$-coordinates.  }
\end{figure}
In figure 2 we plotted the Penrose diagram of our solution. The only difference with the original BTZ Penrose diagram is the location of the lines  $r=\infty$.
The maximal extension is then achieved by glueing together  the regions II, III and IV.
\section{The conformal invariant solution}
We consider now the 4D spacetime (compare with  Eq.(\ref{1}))
\begin{equation}
ds^2=\omega(r)^2\Bigl[-N(r)^2 dt^2+\frac{1}{N(r)^2}dr^2+dz^2+r^2\Bigl(d\varphi+N^\varphi(r)dt\Bigr)^2\Bigr]\label{11},
\end{equation}
where $\omega$ represents the dilaton field. So we can write the metric as $g_{\mu\nu}=\omega^2 \tilde g_{\mu\nu}$, where  $\tilde g_{\mu\nu}$ is now called an "un-physical" spacetime.
The conformal invariant Lagrangian\cite{wald:1984,thooft:2015} is
\begin{eqnarray}
{\cal S}=\int d^4x\sqrt{- \tilde g}\Bigl\{\omega^2 \tilde R+6\partial_\alpha\omega\partial^\alpha\omega+\kappa^2\Lambda\omega^4\Bigr\}\label{12}.
\end{eqnarray}
One can easily proof that this Lagrangian is local  conformal invariant under
\begin{eqnarray}
\tilde g_{\mu\nu}({\bf x})\rightarrow \Omega({\bf x})^2\tilde g_{\mu\nu}({\bf x}), \qquad \omega({\bf x})\rightarrow \frac{1}{\Omega({\bf x})}\omega({\bf x})\label{13}.
\end{eqnarray}
The field equations become
\begin{eqnarray}
\tilde G_{\mu\nu}=\frac{1}{\omega^2}\Bigl(\tilde T_{\mu\nu}^{(\omega)}+\frac{1}{6}\tilde g_{\mu\nu}\Lambda\kappa^2\omega^4\Bigr), \quad
\tilde\nabla^\alpha \partial_\alpha\omega -\frac{1}{6}\tilde R\omega-\frac{1}{9}\Lambda \kappa^2\omega^3=0\label{14},
\end{eqnarray}
with
\begin{eqnarray}
\tilde T_{\mu\nu}^{(\omega)}=\Bigl(\tilde\nabla_\mu\partial_\nu\omega^2-\tilde g_{\mu\nu}\tilde\nabla_\alpha\partial^\alpha\omega^2\Bigr)
-6\Bigl(\partial_\mu \omega\partial_\nu \omega-\frac{1}{2}\tilde g_{\mu\nu}\partial_\alpha \omega\partial^\alpha\omega)\Bigl).\label{15}
\end{eqnarray}
The cosmological constant $\Lambda$ could be ignored from the point of view of naturalness in order to avoid the inconceivable  fine-tuning. Putting $\Lambda$ zero increases the symmetry of the model.
Note that the covariant derivatives are taken with respect to $\tilde g_{\mu\nu}$ and $\tilde R$ is associated to $\tilde g_{\mu\nu}$.
\begin{figure}[h]
\centerline{
\includegraphics[width=4.3cm]{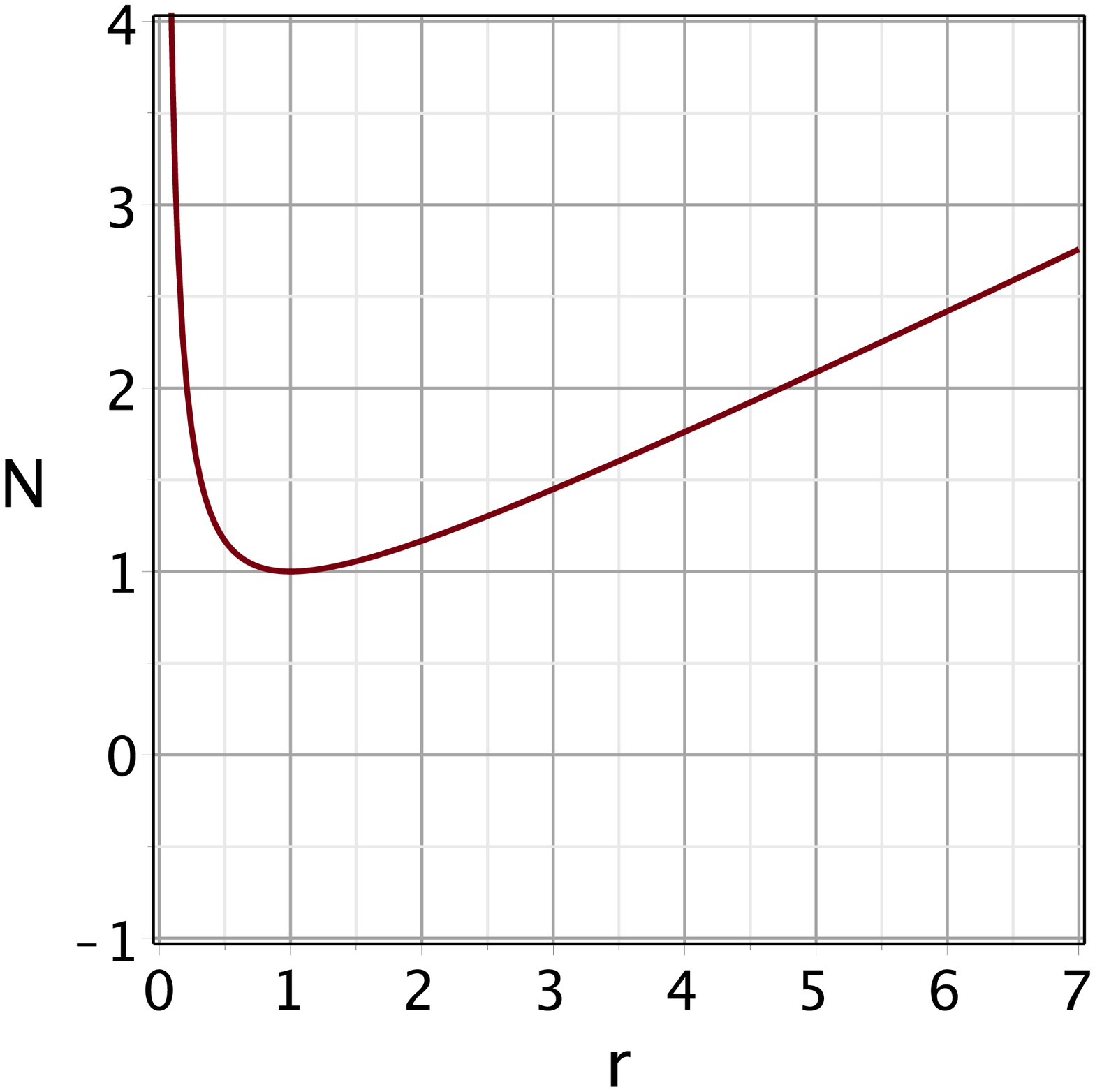}
\includegraphics[width=4.3cm]{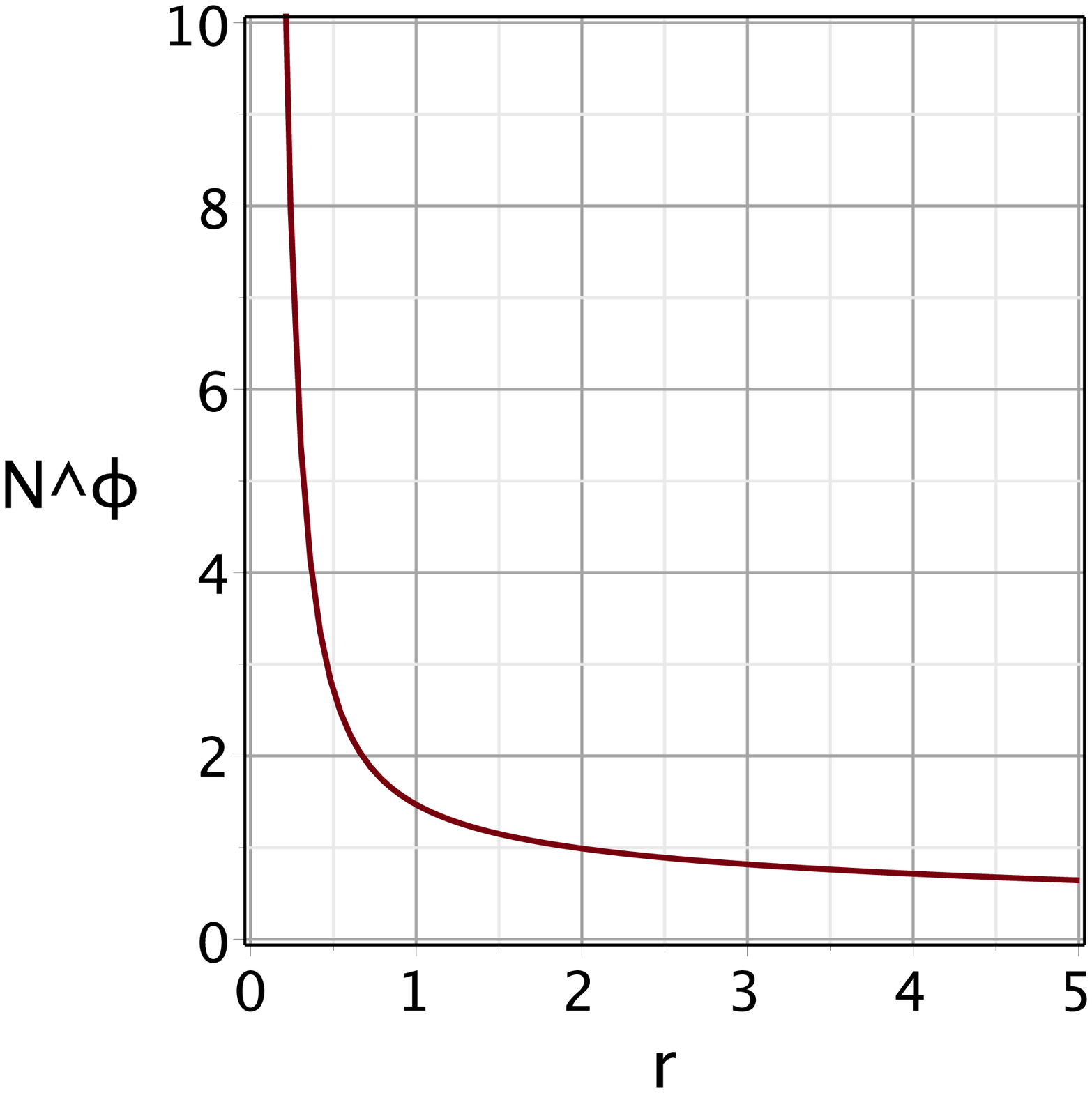}
\includegraphics[width=4.3cm]{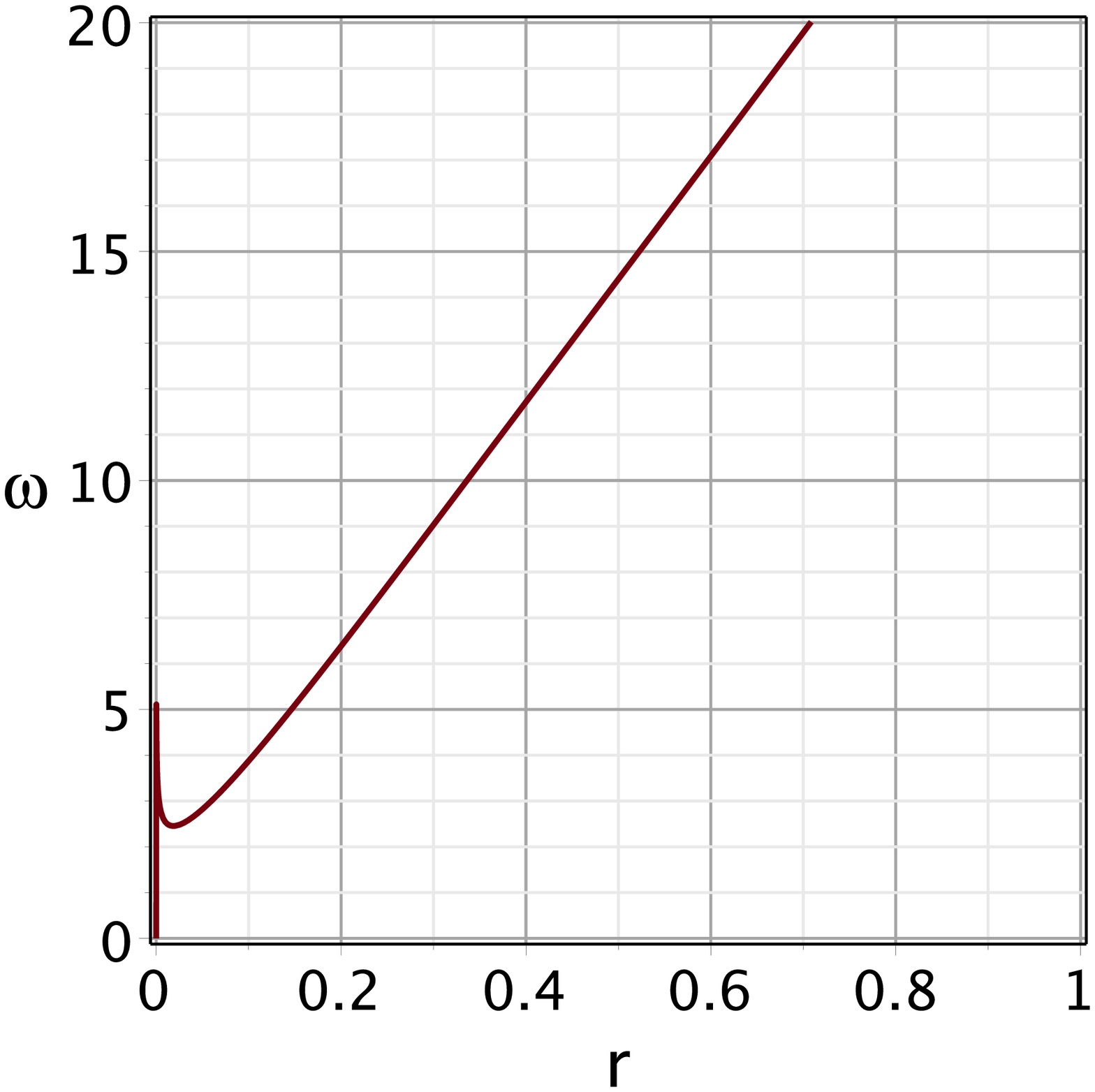}}
\vspace{1.cm}
\caption{Example of a conformal invariant solution of the BTZ 4D spacetime.  }
\end{figure}
The set of differential equations become
\begin{eqnarray}
N''=\frac{1}{3r^2N}\Bigl[r^2(N')^2-3rN N'+2N^2\bigr],\\
N{^\varphi}^{''}=\frac{N{^\varphi}^{'}}{3rN}(7N-4rN'),\\
\omega ''=\frac{2\omega}{9r^2N^2}(2rN'+N)^2.\label{16}
\end{eqnarray}
An exact solution can be found,
\begin{eqnarray}
N=\frac{(c_1r^{4/3}+c_2)^{3/2}}{r},  \quad N^\varphi=c_3+c_4\Bigl(\frac{c_2^2}{4r^{8/3}}+\frac{c_1c_2}{r^{4/3}}-\frac{2}{3}c_1^2\log(r)\Bigr), \cr
\qquad \omega =(c_1r^{4/3}+c_2)^2\Bigl[c_5H_+(r)r^{1/2+1/6\sqrt{17}}+c_6H_-(r)r^{1/2-1/6\sqrt{17}}\Bigr],\label{17}
\end{eqnarray}
with
\begin{equation}
H_{\pm}(r)=Hypergeom\Bigl(\Bigl[\frac{7}{8}\pm\frac{1}{8}\sqrt{17},\frac{25}{8}\pm\frac{1}{8}\sqrt{17}\Bigl],\Bigl[1\pm\frac{1}{4}\sqrt{17}\Bigr],\frac{c_4r^{4/3}}{c_3}\Bigr)\label{18}
\end{equation}
a hyper-geometrical function. The constant $c_1$ and $c_2$ are related to the angular momentum and mass.
By considering the BTZ spacetime as the un-physical metric $\tilde g_{\mu\nu}$ in the conformal invariant setting, we  can then consider  the dilaton as the scale factor.
$\tilde g_{\mu\nu}$ has evidently the horizon at $r=\pm\frac{(c_2)^{3/4}}{c_1}$ and an ergo-circle, while $g_{\mu\nu}$ has no horizon, although it has  the ergo-circle at $r=\pm\frac{(c_2)^{3/4}}{c_1}$. So one could say that the boundary of the object lies at $r=\pm\frac{(c_2)^{3/4}}{c_1}$.
In figure 3 we plotted a typical solution for some values of $c_i$.
It is clear that the solution for $g_{\mu\nu}$ is then regular and does not suffer from the problems encountered in the BTZ solution, such as the up-lifting problem to (3+1) dimensional models and the formation of closed timelike curves and naked singularities.
\subsection{The 3D counterpart solution}
If we disregard the $dz^2$ term, we obtain the solution
\begin{eqnarray}
\omega=\frac{1}{c_1r+c_2},\qquad N^\varphi =c_3+c_4\Bigl[c_1^2\log(r)-\frac{c_2}{2r^2}(4c_1r+c_2)\Bigr],\cr
N=\frac{(c_1r+c2)^4}{\sqrt{10}r}\qquad \qquad \qquad\label{19}
\end{eqnarray}
\begin{figure}[h]
\centerline{
\includegraphics[width=4.3cm]{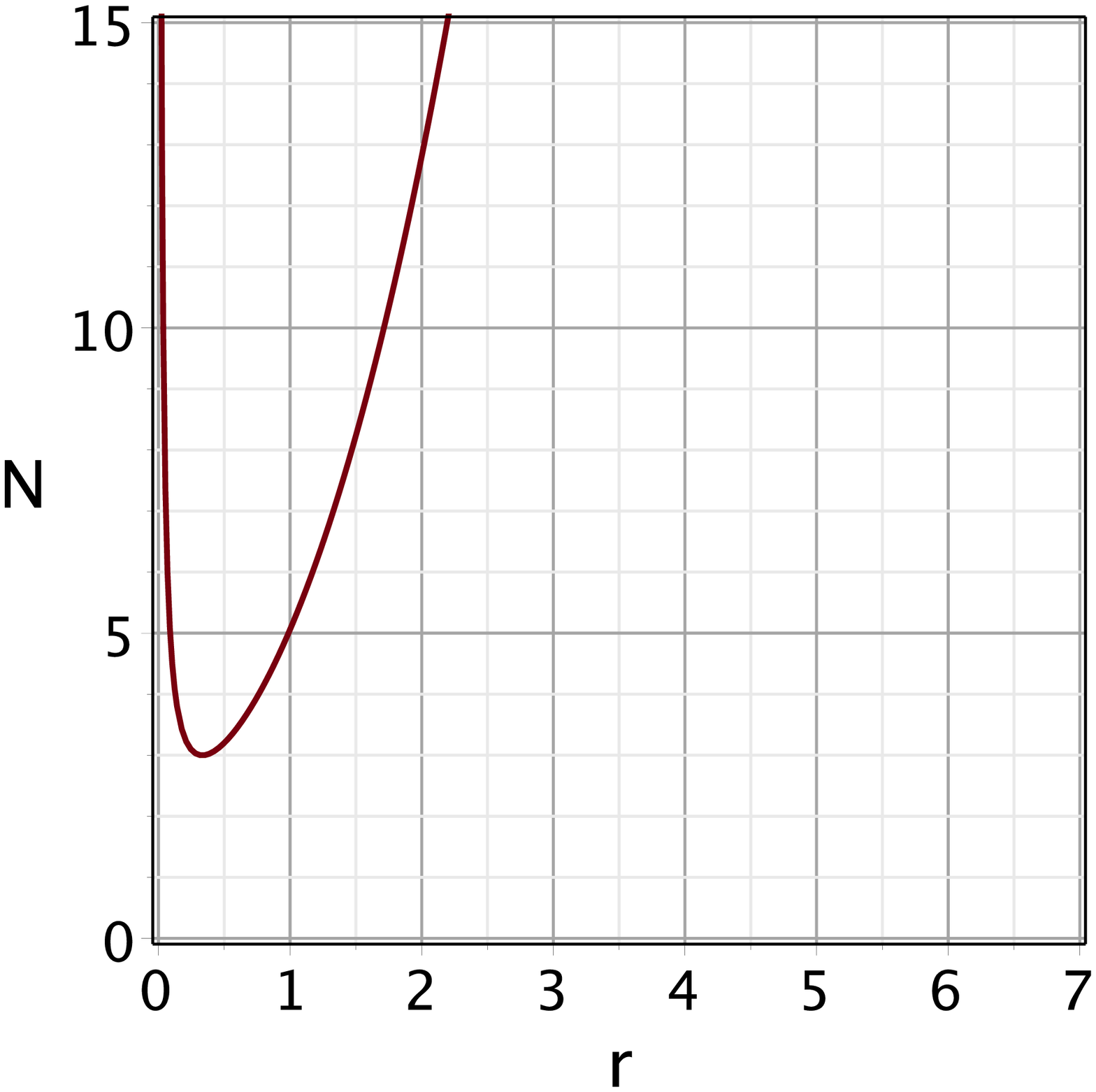}
\includegraphics[width=4.3cm]{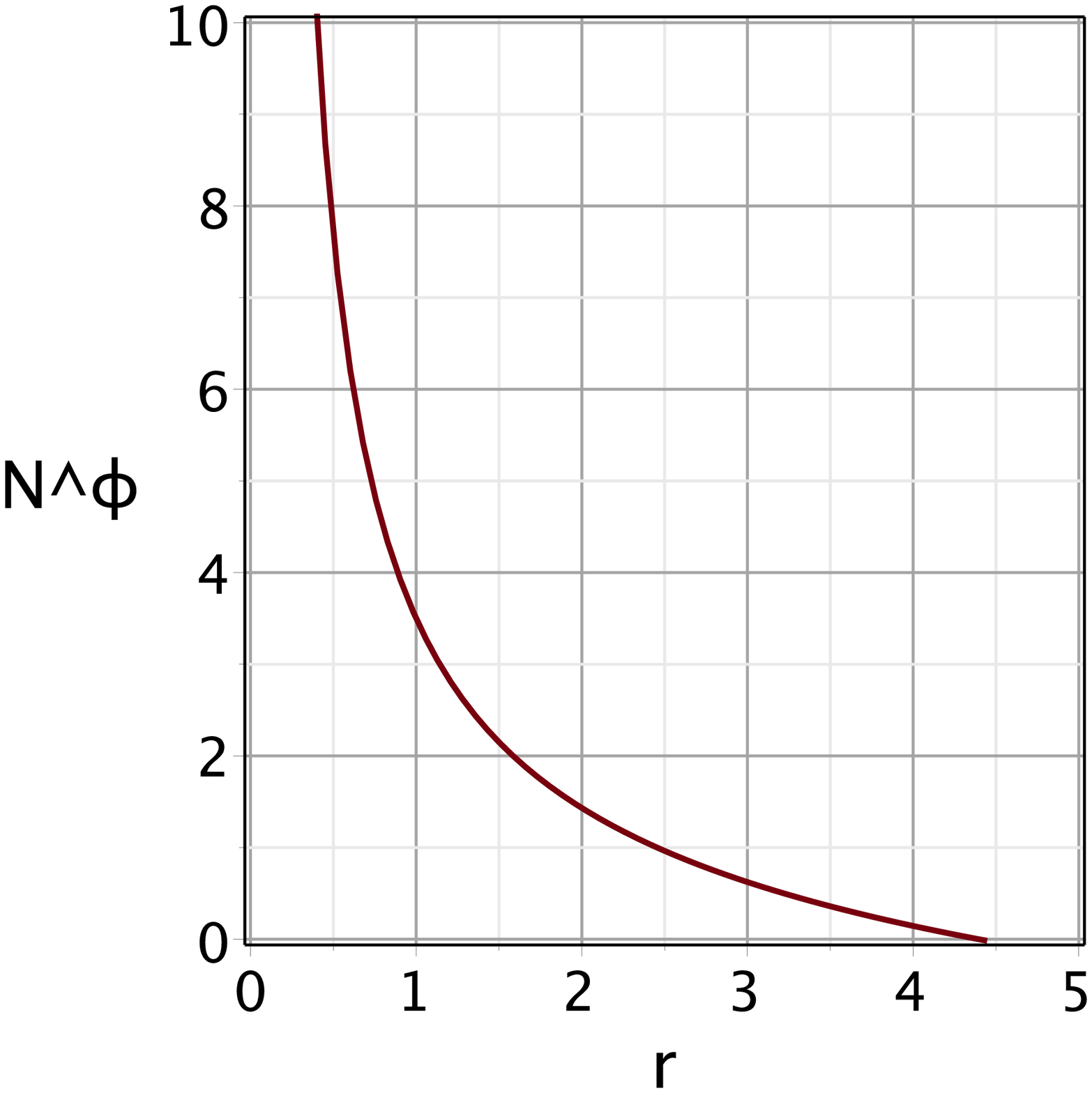}
\includegraphics[width=4.3cm]{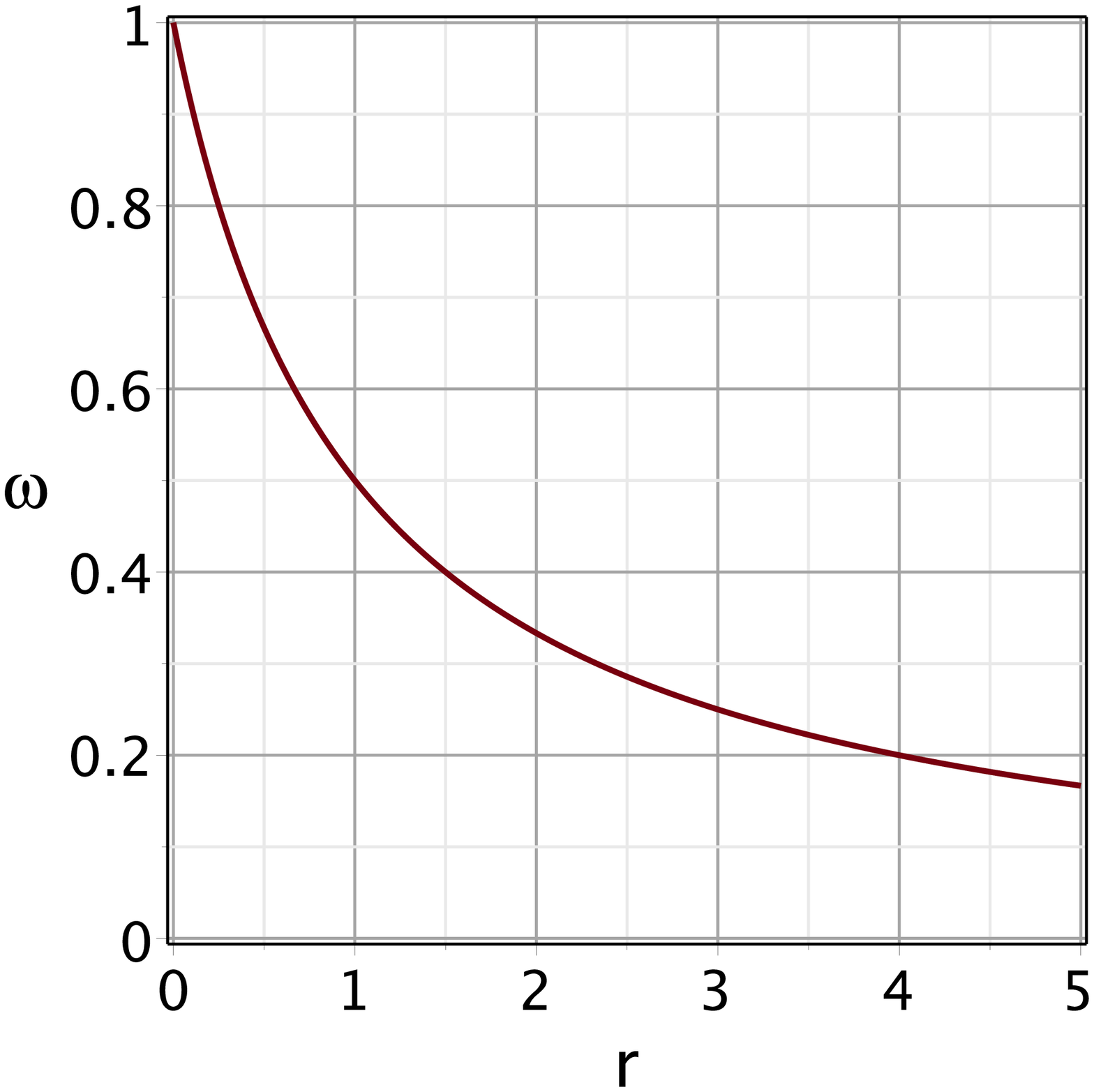}}
\vspace{1.cm}
\caption{Example of a conformal invariant solution of the BTZ in 3D spacetime.  }
\end{figure}
In figure 4 we plotted  a typical solution. If we compare this solution with the 4D counterpart solution of Eq.(\ref{17}), we observe that only the behavior of $\omega$ differs significantly.
This could be explained as follows. Locally at small scales one considers the 3D conform invariant model where an observer experiences $\omega$ as given by Eq.(\ref{19}), while at larger scales, the 4D counterpart model describes a different $\omega$ given by Eq.(\ref{17}).
Another argument in favour of our model concerning the physical acceptability, is the issue of the mass of the object. This will be treated in the next section.
One must keep in mind that the dilaton field must be  treated as a quantum field, when approaching smaller scales\cite{thooft:2015}.  If one incorporates matter fields in the model, then the local conformal invariance will be broken.
These issues will not be further pursued here.

\section{The  spinning cosmic string connection}
In a former study\cite{slagter:2019} we found  an exact Ricci-flat conformal invariant solution of the exterior of a spinning cosmic string of the metric
\begin{equation}
ds^2=\omega(r)^2\Bigl[-(dt-J(r)d\varphi)^2+b(r)^2d\varphi^2+e^{2\mu(r)}(dr^2+dz^2)\Bigr]\label{20}.
\end{equation}
which could be matched on the interior of the U(1) scalar-gauge field solution.
The most important results of the solution were the correct asymptotic behavior of $J(r)$ and the absent of horizons. For the exterior it was found that
\begin{equation}
J(r)=cons.\int\frac{b(r)}{ \omega(r)^2}dr\label{21}
\end{equation}
A comparable relation can be found here in the 4D as well as in the 3D case, i.e.,
\begin{equation}
N^\varphi(r)=cons.\int\frac{1}{r^3\omega(r)^2}dr\label{22}
\end{equation}

So it is conjectured that the new CI BTZ solution can possible be linked to an interior mass of the cosmic string.
In the original BTZ solution, localized matter has no influence on the local geometry of the source free regions and effects only the global spacetime. The asymptotic symmetry group (considered as gauge transformations)
is then applied and boundary conditions are adopted at spatial infinity. After the coordinate transformation (see Eq.(\ref{4})), the "jump" in the coordinate time was related to the angular momentum and the angle deficit with the mass. Only at $r'=0$ there is an obstruction.
In our model, after up-lifting to 4D, we have a boundary determined by the matter fields of the spinning cosmic string. One can easily check that in the case of global strings, with a scalar field $\Phi$ present, the relation between the angular momentum and the dilaton field of Eq.(\ref{21}) changes into\cite{slagter:2019}
\begin{equation}
J(r)=cons.\int\frac{b(r)}{\omega(r)^2+\eta^2\Phi(r)^2}dr\label{23}
\end{equation}
with $\eta$ the vacuum expectation value. Our solution of section 3 suggest that it is the dilaton field that determines the global behavior of the spacetime and not an infinite thin line mass. At very small scales when
$\omega\rightarrow 0$, $J(r)$ remains finite.
Moreover, the angular momentum has already the correct asymptotic behavior. In the up-lifted situation, $\omega$ is a truly scale factor.
It is conjectured that on small scales, $\omega$ plays a fundamental role in describing evaporating black holes\cite{thooft:2009}.
This issue it currently under investigation by the authors.
\section{Conclusion}
A new solution is found for the BTZ spacetime, without a cosmological constant. The solution shows some different features with respect to the standard BTZ solution. A local non-rotating observer has no coordinate angular velocity, i.e., no frame dragging and there is no jump in the radial coordinate (in a suitable coordinate system) when crossing the horizon. The conformal invariant counterpart model shows no horizon and can be related to the spinning point particle solution of the dimensional reduced spinning cosmic string. It is conjectured that the new solution can be linked to the mass of the interior of the spinning cosmic string.


\end{document}